\documentclass[prl,twocolumn,showpacs,nobibnotes,floatfix,superscriptaddress]{revtex4}

\usepackage{graphicx,eucal}

\begin{document}

\title{Pseudo-random operators of the circular ensembles}

\author{Yaakov S. Weinstein}
\thanks{To whom correspondence should be addressed}
\email{weinstei@dave.nrl.navy.mil}
\author{C. Stephen Hellberg}
\email{hellberg@dave.nrl.navy.mil}
\affiliation{Center for Computational Materials Science, Naval Research Laboratory, Washington, DC 20375 \bigskip}

\begin{abstract}
We demonstrate quantum algorithms to implement pseudo-random operators that 
closely reproduce statistical properties of random matrices from the 
three universal classes: unitary, symmetric, and symplectic. Modified 
versions of the algorithms are introduced for the less experimentally 
challenging quantum cellular automata. For implementing pseudo-random 
symplectic operators we provide gate sequences for the unitary 
part of the time-reversal operator.
\end{abstract}


\maketitle

The possibility of manipulating, transferring and storing information in a 
way that preserves quantum coherence has led to a reexamination and extension 
of the postulates of information processing \cite{N}. This field of study, 
known as quantum information processing, can claim as triumphs the discovery 
of quantum algorithms that factor large numbers \cite{Shor}, search 
databases quickly \cite{Grover}, and simulate quantum systems efficiently 
\cite{L1}. In addition, powerful communication and cryptographic protocols 
have been suggested based on the laws of quantum mechanics \cite{N}. 

Random number generation is a basic component of classical information theory.
Their quantum counterparts, random quantum states and operators, likewise
play a vital role in quantum information theory. Quantum communication 
protocols utilizing randomness include saturation of the 
classical communication capacity of a noisy quantum channel by random 
states \cite{Seth2} and superdense coding of quantum states via random 
operators \cite{Aram}. Quantum computing protocols facilitated by random 
unitaries include quantum process tomography via a random operator fidelity 
decay experiment to identify types and strengths of noise generators 
\cite{RM}. In addition, the amount of multi-partite entanglement in random 
states approaches the maximum at a rate exponential with the number of qubits 
in the system \cite{Scott}.

Random quantum states, generated by applying a random operator 
to computational basis state, can also be used for unbiased sampling. When 
testing an algorithm, such as quantum teleportation, or a communications 
scheme it is desirable to insure success for all possible quantum 
states. This can be done via quantum process tomography, however this is 
extremely inefficient \cite{QPT}. Rather, one could test the likelihood of 
success with states drawn in an unbiased manner from the space of all 
quantum states, similar to sampling statistics in other contexts.

To capitalize on the above uses of random states and operators, it is 
necessary to efficiently implement random matrices on a quantum computer.
This would appear to be a daunting task considering that the number of 
independent variables in a given random operator grows exponentially with 
matrix dimension. Nevertheless, pseudo-random operators suggest that it 
may be possible to efficiently reproduce statistical properties of randomness 
on a quantum computer \cite{RM}. In this paper we extend the algorithm of 
Ref. \cite{RM} to produce pseudo-random operators from the universal 
random matrix classes with time-reversal symmetry. 

Pseudo-random operators from the universal classes with time-reversal 
symmetry may help a quantum computer simulate systems with time-reversal 
symmetries. These systems are especially important in the areas of 
quantum chaos \cite{Haake} and decoherence \cite{Gorin}, where the physical 
system or environment to be studied tend to be modeled with time-reversal 
symmetry. We note that models of decoherence, specifically with random 
classical fields, have already been implemented on a nuclear magnetic 
resonance quantum information processor \cite{Grum}. 

In addition, we extend our recent work \cite{QCARM} and show that random 
operators from all three universal classes can be implemented on the less 
experimentally challenging quantum cellular automata (QCA) architecture. 
This further demonstrates the usefulness of a QCA in the study of complex 
quantum evolution. For both architectures we show how to implement the 
unitary part of the time-reversal operator on a quantum computer which, 
for the QCA case, requires identifying a suitable, non-standard 
form of the time-reversal operator.

Random matrices were first introduced by Wigner to describe the energy level 
spacings of large nuclei \cite{Wigner}. Since then, random matrices 
have functioned as a universal model for a host of complex systems 
ranging from quantum dots to field theory \cite{RMT}. The circular ensembles 
of unitary matrices were introduced by Dyson \cite{Dyson} as alternatives 
to the Gaussian ensembles of Hermitian matrices \cite{Wigner,Mehta}. 
The three circular ensembles are the circular unitary ensemble (CUE) of 
arbitrary unitary matrices, appropriate for modeling systems without time 
reversal symmetry, the circular orthogonal ensemble (COE) of symmetric 
unitary matrices, appropriate for systems having time-reversal invariance and 
integral spin or rotational symmetry, and the circular symplectic ensemble 
(CSE) of self-dual unitary quaternion matrices, appropriate for systems 
with time-reversal invariance, half-integer spin, and no rotational 
symmetries. Each universality class has properties unique unto itself. 
For example, the degree of level repulsion (the rate 
of change of nearest neighbor eigenangle spacings as the spacing goes 
to zero) is $P(s) \sim s$ for COE, $P(s) \sim s^2$ for CUE, and 
$P(s) \sim s^4$ for CSE, where $s$ is the nearest neighbor eigenangle 
spacing. Additionally, the distribution of eigenvector component amplitudes 
follow the $\chi^2_{\nu}$ distribution 
\cite{Zyc1}
\begin{equation} 
P_{\nu}(y) = \frac{\nu/2^{(\nu/2)}}{\Gamma(\nu/2)\langle y \rangle}\Big(\frac{y}{\langle y \rangle}\Big)^{\nu/2-1}\exp\Big(\frac{\nu y}{2\langle y \rangle}\Big),
\end{equation}
where $y$ is the eigenvector component amplitude, and $\langle y \rangle$ is 
the mean value of $y$. The number of degrees of freedom, $\nu$, is 1 for the 
orthogonal ensemble, 2 for the unitary ensemble, and 4 for the symplectic 
ensemble. 

\begin{figure}
\includegraphics[height=5.8cm, width=8cm]{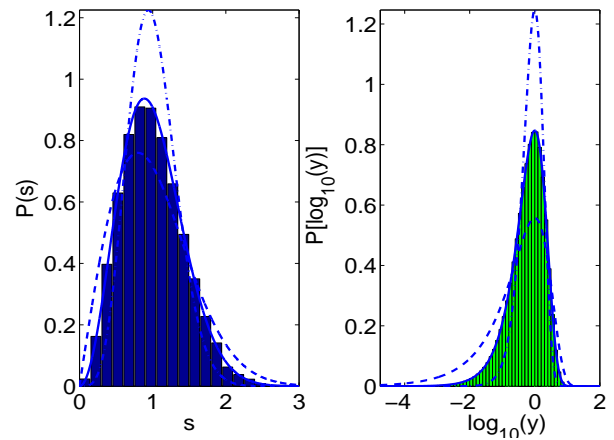}
\caption{\label{CUE}
Distributions of nearest neighbor eigenangle spacings, $s$ (left), and 
eigenvector component amplitudes, $y$ (right), for 100 realizations of 
8-qubit, 60 iteration pseudo-random CUE operators compared to that 
expected for random unitaries of COE (dash), CUE (solid), and CSE (dash-dot). 
The nearest neighbor eigenangle spacing distribution compares extremely well 
with the expected distribution $P_{CUE}(s) = (32s^2/\pi^2)\exp(-4s^2/\pi)$ and
the operators' eigenvector component amplitude distribution almost exactly 
follows $P_{CUE}(y) = \exp(-y)$, which is appropriate when 
$\langle y \rangle = 1$ and in the limit $N\rightarrow \infty$. 
}
\end{figure}

The algorithm introduced in \cite{RM} to produce pseudo-random operators of 
arbitrary unitaries, CUE, consists of $m$ iterations of the $n$ qubit 
gate: apply a random SU(2) rotation to each qubit, then evolve the system via 
all nearest neighbor couplings \cite{RM}. A random SU(2) rotation on qubit 
$j$ of iteration $i$ is defined as \cite{Zyc1}
\begin{eqnarray}
R(\theta^j_i,\phi^j_i, \psi^j_i) &=&
\left(
\begin{array}{cc}
e^{i\phi^j_i}\cos\theta^j_i & e^{i\psi^j_i}\sin\theta^j_i \\ 
-e^{-i\psi^j_i}\sin\theta^j_i &  e^{-i\phi^j_i}\cos\theta^j_i \\
\end{array}
\right), 
\end{eqnarray}
where the angles $\phi^j_i$, and $\psi^j_i$ are drawn uniformly from the 
intervals
\begin{equation}
0\leq \phi^k_i \leq 2\pi \;\;\;\;\;\; 0\leq \psi^k_i \leq 2\pi,
\end{equation}
and $\theta^j_i = \sin^{-1}({\xi^j_i}^{1/2})$ where $\xi^j_i$ 
is drawn uniformly from 0 to 1. The nearest neighbor coupling operator 
at every iteration is
\begin{equation}
\label{nnc}
U_{nnc} = \exp\left(i(\pi/4)\sum^{n-1}_{j=1}\sigma_z^j\otimes\sigma_z^{j+1}\right),
\end{equation}
where $\sigma_z^j$ is the $z$-direction Pauli spin operator.
The random rotations are different for each qubit and each iteration, but the 
coupling constant is always $\pi/4$ to maximize entanglement. After the $m$ 
iterations, a final set of random rotations is applied. This algorithm has 
been shown, for up to 10 qubits, to implement operators with statistical 
properties extremely close to those expected of CUE \cite{RM} with 
relatively few iterations \cite{REQC}, Fig. \ref{CUE}. 

\begin{figure}
\includegraphics[height=5.8cm, width=8cm]{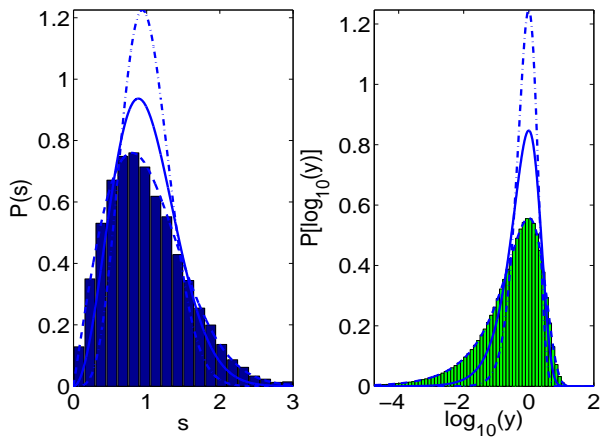}
\caption{\label{COE}
Distributions of nearest neighbor eigenangle spacings, $s$ (left), and 
eigenvector component amplitudes, $y$ (right), for 
100 realizations of 8-qubit, 60 iteration pseudo-random COE operators. 
The nearest neighbor eigenangle distribution compares 
very well with the distribution $P_{COE}(s) = (\pi s/2)\exp(-\pi s^2/4)$ 
and the eigenvector component amplitude distribution almost exactly follows 
$P_{COE}(y) = (1/\sqrt{2\pi y})\exp(-y/2)$. 
}
\end{figure}

Random operators from the other two universal classes can be 
constructed from CUE operators. The goal of this paper is to demonstrate 
that these constructions allow for simple modifications of the pseudo-random 
algorithm described above to generate pseudo-random operators from these 
classes. To draw a matrix, $U_{COE}$ from the space of all COE simply draw 
$U_{CUE}$, from CUE and multiply it by its transpose \cite{Mehta,Zyc1}
\begin{equation} 
U_{COE} = U_{CUE}^TU_{CUE}.
\end{equation}
Pseudo-random generation of such an operator is readily done. First, 
implement the CUE matrix $U_{CUE}$ as above, retaining in memory (quantum or
classical) the values of the $3n(m+1)+1$ independent variables necessary to 
implement the operator, 3 for each rotation of $n$ qubits for $m+1$ 
rotations and 1 for the coupling strength. Next, implement the transpose 
of the operator by applying the transpose of each specific operation 
in reverse order of the original $U_{CUE}$ generation. 
To implement the transpose of the rotations apply the transpose of each 
individual qubit rotations. The transpose of the coupling operation is the 
same coupling $U_{nnc}^T = U_{nnc}$. In Fig. \ref{COE} we demonstrate that 
for 8 qubits and 60 iterations the pseudo-random COE matrices generated in 
this way fulfill statistics of randomness by comparing the distributions 
$P(s)$ and $P(y)$ to that expected for COE.

We now turn to the symplectic ensemble, representing systems 
with half-integer spin that are invariant under time-reversal. Following 
Mehta \cite{Mehta} we define the anti-unitary time-reversal operator 
$T = ZC$, where $C$ takes the complex conjugate and $Z$ 
is unitary. The symplectic ensemble is characterized by an anti-symmetric 
$Z$, i.e. $ZZ^* = -1$ where $^*$ means non-Hermitian conjugate. We choose the 
representation such that $Z$ is written 
$I_1\otimes I_2\otimes \dots \otimes I_{n-1}\otimes z_n$ 
where $I_j$ is the two-dimensional identity matrix and 
\begin{eqnarray}
z_j &=&
\left(
\begin{array}{cc}
0 & -1 \\
1 & 0 \\
\end{array}
\right).
\end{eqnarray}
In this way, a symplectic unitary is defined by 
\begin{equation}
U_{CSE}^R \equiv  -ZU_{CSE}^TZ = U_{CSE}.
\end{equation}
and $U_{CSE}Z$ is anti-symmetric unitary. 


As with the COE operators, drawing an operator from CSE can be done via CUE 
operators: draw $U_{CUE}$ and multiply by its time-reversal 
\cite{Zyc2,Mehta}
\begin{equation}
U_{CSE} = U_{CUE}^RU_{CUE},
\end{equation}
where $U_{CUE}^R = -ZU_{CUE}^TZ$. Using this construction pseudo-random CSE 
operators can be generated as follows: run the pseudo-random operator 
algorithm to implement $U_{CUE}$, apply $Z$ via two rotations of the 
least significant qubit $z = \exp(-i(\pi/2)\sigma_z)\exp(-i(\pi/2)\sigma_x)$, 
where the $\sigma_i$ are the Pauli matrices, $U_{CUE}^T$ is  
implemented as explained in the COE case, apply $Z$. The negative sign 
is a global phase. 

\begin{figure}
\includegraphics[height=5.8cm, width=8cm]{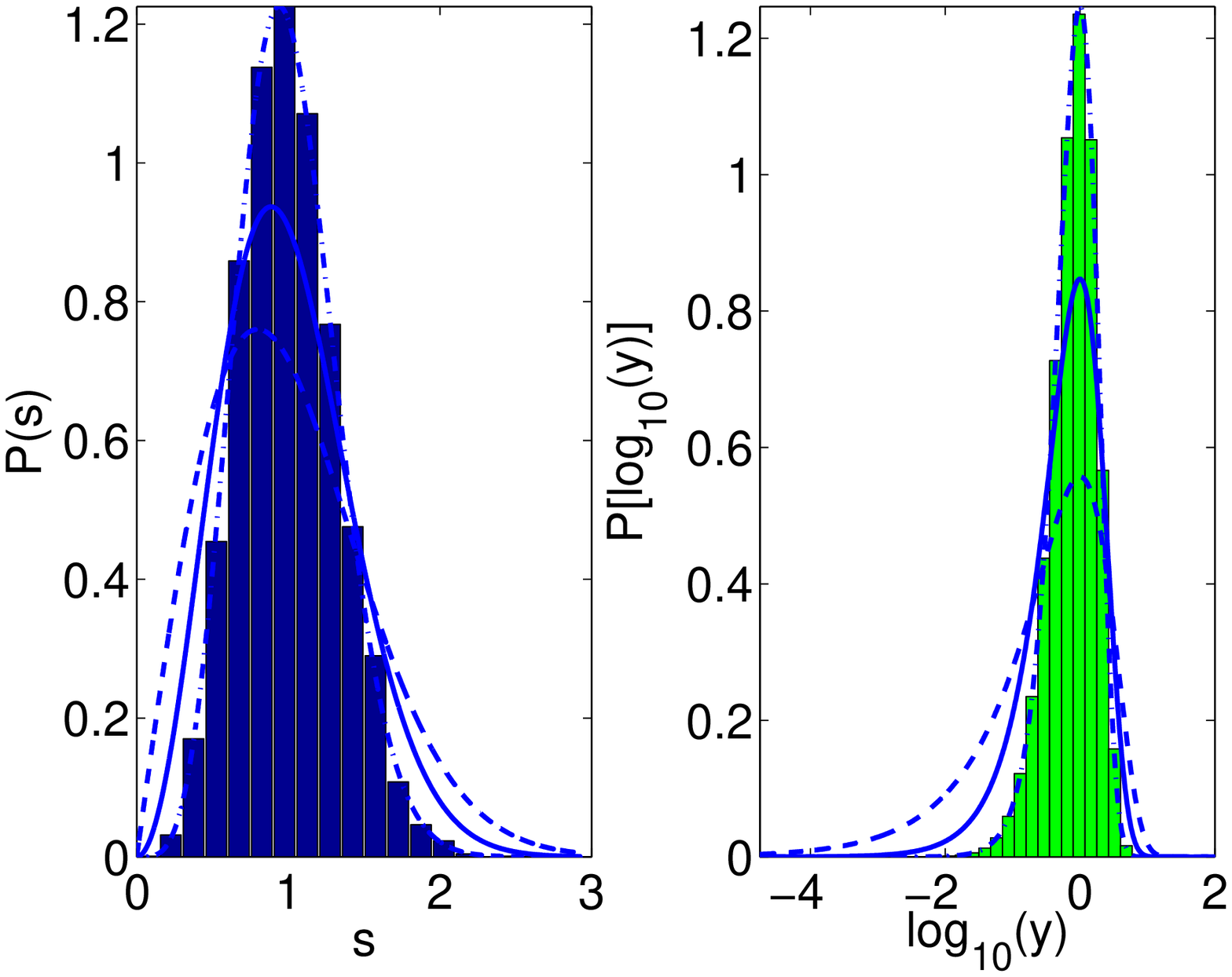}
\caption{\label{CSE}
Distributions of nearest neighbor eigenangles, $s$ (left), and eigenvector 
component amplitudes, $y$ (right), for 
100 realizations of 8-qubit, 60 iteration pseudo-random CSE operators. 
The nearest neighbor eigenangle distribution compares very well with the 
distribution $P_{CSE}(s) = (64/9\pi)^3s^4\exp(-64s^2/9\pi)$ and
the eigenvector component amplitude distribution almost exactly follows 
$P_{CSE}(y) = 4y\exp(-2y)$.
}
\end{figure}

Figure \ref{CSE} shows the eigenangle and eigenvector element distributions
for CSE pseudo-random operators. We note that matrices of the CSE 
exhibit Kramers' degeneracy so we digress to explain how the above
distributions are determined. Kramers' degeneracy allows the following basis 
choice for CSE matrices \cite{Haake}
\begin{equation}
|1\rangle, T|1\rangle, |2\rangle, T|1\rangle, \dots |N/2\rangle, T|N/2\rangle, 
\end{equation}
where $T$ is the time-reversal operator. The nearest neighbor eigenangle 
distribution uses only one of each degenerate eigenangle. The eigenvectors 
corresponding to the degenerate eigenangles can be written as 
\begin{eqnarray}
|e_1\rangle &=& c_1|1\rangle+\tilde{c}_1T|1\rangle+c_2|2\rangle+\tilde{c}_2T|2\rangle\dots \nonumber\\
T|e_1\rangle &=& -\tilde{c}^*_1|1\rangle+c^*_1T|1\rangle-\tilde{c}^*_2|2\rangle+c^*_2T|2\rangle\dots
\end{eqnarray}
Any given diagonalization code will not necessarily output the above form for
the two eigenvectors of a degenerate eigenvalue, but superpositions 
of the two. Thus, as an invariant quantity to characterize the eigenvectors 
we use $y = |c_1|^2+|\tilde{c}_1|^2$ \cite{Haake}. Using the above 
procedures, the distribution of nearest neighbor eigenangle and eigenvector 
components for the generated pseudo-random CSE operators are those shown in 
Fig. \ref{CSE}.

Classical cellular automata have been used to simulate many complex classical 
systems from crystal growth to fluid flow \cite{Wolf}. Thus, one may expect 
that quantum cellular automata (QCA) can be used to model complex quantum 
systems. Ref. \cite{QCARM} demonstrates the implementation of CUE 
pseudo-random operators. Here we extend that work to the other two random 
matrix classes.

A QCA system is devised of $k$ species of qubits in which all qubits of a 
species are addressed simultaneously and equivalently. Experimental 
flexibility is a primary motivation to explore implementations via QCA. 
Removing the need for localized external Hamiltonians can greatly ease 
hardware specifications for actual implementations of quantum information 
processing. A number of works have been devoted to exploring the 
universality of QCA architectures \cite{Seth,W,Benj} but, despite the greater 
experimental ease of QCA, relatively little work has been done to 
exploit the uniqueness of the QCA architecture \cite{Bren,QCARM}.

Previous work has shown that the pseudo-random algorithm applied
to a one-species QCA chain, such that all qubits undergo the same rotations, 
yields operators with eigenvalue and eigenvector distributions appropriate
for CUE-type operators with mirror symmetries \cite{QCARM}. The use of a two
species QCA with alternating qubit species or the change of one nearest 
neighbor coupling constant (say from $\pi/4$ to $\pi/5$) is sufficient to 
break this symmetry.

\begin{figure}
\includegraphics[height=5.8cm, width=8cm]{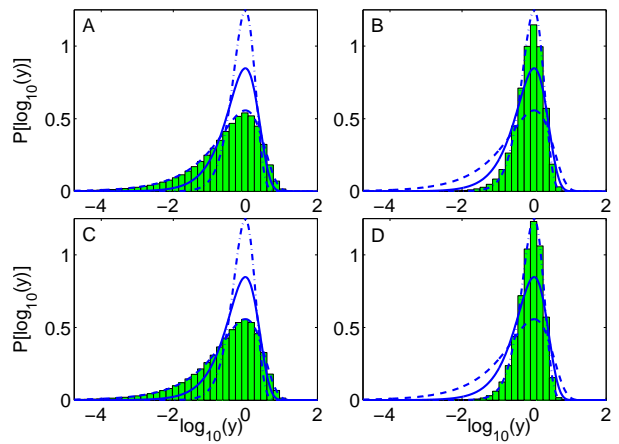}
\caption{\label{QCAvecs}
Distributions of eigenvector component amplitudes, $y$, for quantum cellular
automata (QCA) based operators: (A) 100 realizations of 8-qubit one-specie 
COE, (B) 200 realizations of 7-qubit one-specie CSE, (C) 100 realizations of 
8-qubit two-species COE, (D) 200 realizations of 7-qubit two-species CSE. 
All distributions are for $m = 40$ iterations. The distributions of the 
two-species QCA operators are indistinguishable from those of random COE and 
CSE matrices. However, the one-specie QCA operator distributions deviate 
from the random distributions due to mirror symmetry. We note that this 
symmetry would be broken in any actual experimental implementation due to 
unequal nearest-neighbor couplings.} 
\end{figure}

For a QCA COE pseudo-random operator, one applies the pseudo-random operator 
algorithm, with all qubits of a species undergoing the same rotation, 
followed by its transpose, just as in the circuit architecture. The 
eigenvector component amplitude distribution for 8-qubit, one- and 
two-species COE operators is shown in fig. \ref{QCAvecs}. 

To generate CSE pseudo-random operators requires the implementation of $Z$ 
which above was done by individually address the least-significant qubit. 
This operation is illegal on a QCA system. Thus, we must find an appropriate 
non-standard representation of $Z$ which allows all qubits of a species
to be addressed equivalently. 

To find a representation of $Z$ amenable to a QCA implementation, we recall 
that for a symplectic matrix, $U_{CSE}$, the matrix $A = U_{CSE}Z$ 
is antisymmetric unitary. For every antisymmetric unitary matrix there exists 
a unitary matrix $W$ such that $A = WZW^T$ \cite{Mehta}.
We define our modified operator as $Z' = VZV^T$. 
Since $Z'Z'^* = -1$, by definition of a symplectic matrix, 
$(V^*)^TV = V(V^*)^T = \pm 1$. Thus, $V$ must be a symmetric or 
antisymmetric unitary. We can then define the antisymmetric unitary 
$A'= U_{CSE}Z'= W'Z'W'^T$ and, following Mehta 
\cite{Mehta}, we choose the unitary $U_{CUE} = (Z'W')^T$ 
and generate symplectic matrices via $U_{CSE} = -Z'U_{CUE}^TZ'U$. An example 
of a symmetric unitary operator, $V$, that allows the generation of 
CSE operators in the above fashion is the swap gate. This should not be 
surprising as the ordering of qubits is completely arbitrary. An operator 
$Z'$ that is appropriate for our purposes is the rotation $z$ applied to an 
odd number of qubits. Using this form of $Z'$, pseudo-random symplectic 
operators (with mirror symmetry if all coupling constants are equal) 
can be implemented on a one-specie QCA if there are an odd number of 
qubits. Similarly, two-species QCA implementations of pseudo-random 
symplectic operators can be achieved if one of the species consists 
an odd number of qubits. The eigenvector component amplitude distribution of 
7-qubit one- and two-species QCA operators is shown in Fig. \ref{QCAvecs}.

In conclusion, we have demonstrated quantum algorithms for pseudo-random 
operators from the COE and CSE universal classes. As with the original CUE
pseudo-random operators \cite{RM}, we provide evidence that suggests that 
these operators may be able to fulfill statistical properties of these 
ensembles with an efficient number of gates. Efficient performance of such 
operators could be useful in simulating various complex quantum systems. 
Similar operators can also be implemented using the less experimentally 
demanding QCA. This is a further demonstration that a QCA system can be a 
useful tool in the study randomness.  

\acknowledgments
The authors would like to thank F. Haake for clarification of the CSE 
eigenvector component statistics and Al. L. Efros for insightful comments. 
The authors acknowledge support from the DARPA QuIST (MIPR 02 N699-00) 
program. Y.S.W. acknowledges the support of the National Research Council 
Research Associateship Program through the Naval Research Laboratory. 
Computations were performed at the ASC DoD Major Shared Resource Center.

\end{document}